\documentclass[prd,12pt]{article}
\pdfoutput=1

\usepackage{tikz}
\usepackage{amsmath,amssymb,graphicx,multirow,xspace,slashed,array,booktabs}
\usepackage[colorlinks=true,urlcolor=blue,anchorcolor=blue,citecolor=blue,filecolor=blue,linkcolor=blue,menucolor=blue,pagecolor=blue]{hyperref}
\usepackage[compress,numbers]{natbib}
\usepackage{subfigure, placeins}
\usepackage{booktabs}
\usepackage{blindtext, rotating}
\usepackage{afterpage}
\usepackage{enumitem}
\usepackage{ marvosym }
\usepackage{authblk} 
\usepackage{verbatim}
\usepackage{soul} 
\usepackage[normalem]{ulem}
\usepackage{pifont}
\usepackage{booktabs}
\usepackage{cancel}

\usepackage{floatrow}
\newfloatcommand{capbtabbox}{table}[][\FBwidth]

\usepackage[font=footnotesize,labelfont=bf]{caption}

\usepackage{lineno}

\allowdisplaybreaks

\long\def\symbolfootnote[#1]#2{\begingroup%
\def\thefootnote{\fnsymbol{footnote}}\footnote[#1]{#2}\endgroup}

\allowdisplaybreaks

\makeatletter
	
	\@addtoreset{equation}{section}
\makeatother

\newcommand{\newc}{\newcommand}
\newc{\gsim}{\lower.7ex\hbox{$\;\stackrel{\textstyle>}{\sim}\;$}}
\newc{\lsim}{\lower.7ex\hbox{$\;\stackrel{\textstyle<}{\sim}\;$}}
\newc{\gev}{\,{\rm GeV}}
\newc{\mev}{\,{\rm MeV}}
\newc{\ev}{\,{\rm eV}}
\newc{\kev}{\,{\rm keV}}
\newc{\tev}{\,{\rm TeV}}
\newc{\MHT}{$H_T^{\text{miss}}$}
\newc{\MET}{$\slashed{E}_T$}
\newc{\MTT}{$M_{T2}$}

\def\ln{\mathop{\rm ln}}

\newc{\mz}{M_Z}
\newc{\mpl}{M_*}
\newc{\mw}{m_{\rm weak}}
\newc{\nr}[1]{N^c_R{}_{#1}}

\def\beq{\begin{equation}}
\def\eeq{\end{equation}}
\newcommand{\bea}{\begin{eqnarray}\begin{aligned}}
\newcommand{\eea}{\end{aligned}\end{eqnarray}}
\def\bitem{\begin{itemize}}
\def\eitem{\end{itemize}}

\begin{document}
\baselineskip 0.6cm

\begin{titlepage}

\vspace*{-0.5cm}

\thispagestyle{empty}

\begin{center}

\vskip 0.5cm

{\LARGE \bf
High Quality Axion \\[1ex] via a Doubly Composite Dynamics
}

\vskip 1cm

\vskip 1.0cm
{\large Seung J. Lee$^1$, Yuichiro Nakai$^2$ and Motoo Suzuki$^2$}
\vskip 0.5cm
{\it
$^1$Department of Physics, Korea University, Seoul 136-713, Korea\\
$^2$Tsung-Dao Lee Institute and School of Physics and Astronomy, \\Shanghai
Jiao Tong University, 800 Dongchuan Road, Shanghai, 200240 China \\}
\vskip 1.0cm

\end{center}

\vskip 0.5cm

\begin{abstract}

We explore a new framework that furnishes a mechanism to simultaneously address the electroweak naturalness problem
and 
the axion high quality problem.
The framework is based on a doubly composite dynamics where the second confinement takes place after the CFT encounters the first confinement and the theory flows into another conformal fixed point.
For a calculable example, we present a holographic dual description of the 4D model via a warped extra dimension model with three 3-branes.
While the hierarchy problem is taken cared of by the localization of the Higgs fields on the TeV brane just as in the original Randall-Sundrum model,
the Peccei-Quinn (PQ) symmetry is realized as a gauge symmetry in the bulk of the extra dimension to solve the axion quality problem.
We introduce a 5D scalar field whose potential at the intermediate brane drives spontaneous breaking of the PQ symmetry.
Then, the PQ breaking scale is given by the scale of the intermediate brane and is naturally small compared to the Planck scale.
The axion bulk profile is significantly suppressed around the UV brane, which protects the axion from
gravitational violations of the PQ symmetry on the UV brane.
Our model genuinely predicts the existence of the Kaluza-Klein excitations of the QCD axion at around the TeV scale
and relatively light extra Higgs bosons.

\end{abstract}

\flushbottom

\end{titlepage}

\section{Introduction}\label{intro}

Nature often shows a hierarchical structure of compositeness.
For example, hadrons are made of quarks, nuclei are composites of protons and neutrons, atoms are made of nuclei and electrons, and so on.
It is natural to imagine that such a hierarchical structure of compositeness continues at energy scales higher than the TeV scale. 
In fact, the Standard Model (SM) cannot be a good effective theory beyond the TeV scale because
the scale of the electroweak symmetry breaking (EWSB) is highly sensitive to a UV cutoff scale.
Emergence of composite nature of the SM Higgs field near the TeV scale is an attractive possibility
(for a review of composite Higgs models, see ref.~\cite{Bellazzini:2014yua}).
The similar naturalness issue arises when we consider the axion solution \cite{Peccei:1977hh,Weinberg:1977ma,Wilczek:1977pj}
to the strong CP problem~\cite{Baker:2006ts,Pendlebury:2015lrz}.
Astrophysical and cosmological observations put constraints on the scale of spontaneous $U(1)_{\rm PQ}$ breaking,
$10^8 \, {\rm GeV} \lesssim f_a \lesssim 10^{12} \, \rm GeV$, which are hierarchically smaller than the Planck scale.
To be worse, for the axion to cancel the QCD vacuum angle $\bar{\theta}$ at the potential minimum,
the  $U(1)_{\rm PQ}$ symmetry must be preserved to an extraordinarily high degree,
in the presence of 
non-perturbative QCD effects.
However, such a global symmetry is not respected by quantum gravity effects.
We naturally expect $U(1)_{\rm PQ}$-violating interactions at the Planck scale
\cite{Holman:1992us,Kamionkowski:1992mf,Barr:1992qq,Ghigna:1992iv,Carpenter:2009zs}
which are dangerous if the axion is
an elementary particle.
An attractive way to solve this axion quality problem is to make the axion composite
\cite{Kim:1984pt,Choi:1985cb,Randall:1992ut,Redi:2016esr,DiLuzio:2017tjx,Lillard:2017cwx,Lillard:2018fdt,Gavela:2018paw,Lee:2018yak}
and realize the $U(1)_{\rm PQ}$ as an emergent symmetry at low-energies. 
Therefore, it is
of highly interest
to consider a possibility that multiple composite dynamics addresses
the naturalness issues for the scales of the EWSB and the $U(1)_{\rm PQ}$ breaking and also realizes a high-quality axion.

Since composite dynamics involves strong interactions, its analysis is not tractable in most cases.
An easier way to deal with 4D composite dynamics is to consider the AdS/CFT correspondence
\cite{Maldacena:1997re,Gubser:1998bc,Witten:1998qj} and
work in the holographic 5D picture.
The Randall-Sundrum (RS) model of a warped extra dimension~\cite{Randall:1999ee}
is understood as the 5D dual picture
of a nearly-conformal strongly-coupled 4D field theory
\cite{ArkaniHamed:2000ds,Rattazzi:2000hs}.
The existence of the IR brane in the RS model corresponds to spontaneous breaking of the conformal symmetry via confinement.
Then, a multi-layer structure of compositeness can be described by a warped extra dimension model with multiple 3-branes.
For example, a 5D model with three 3-branes corresponds to a 4D doubly composite dynamics 
where the second confinement takes place after the CFT encounters the first confinement and the theory flows into another conformal fixed point.
Such multi-brane models have been considered in
\cite{Lykken:1999nb,Hatanaka:1999ac,Kogan:1999wc,Gregory:2000jc,Kogan:2000cv,Mouslopoulos:2001uc,Kogan:2000xc},
originally motivated by cosmology, and properties of a bulk matter field were discussed in
\cite{Mouslopoulos:2001uc,Kogan:2001wp}.
Various phenomenological applications have been explored
\cite{Oda:1999di,Oda:1999be,Dvali:2000ha,Moreau:2004qe,Agashe:2016rle,Agashe:2016kfr,Csaki:2016kqr,Cai:2021nmk}.
And an interesting AdS/CFT discussion with multiple branes can be found in~\cite{Agashe:2016rle}.
In multi-brane models, there exist multiple radion degrees of freedom corresponding to intervals among multiple 3-branes.
These radions are massless without any stabilization mechanism.
Recently, in ref.~\cite{Lee:2021wau} (see also refs.~\cite{Pilo:2000et,Kogan:2001qx,Choudhury:2000wc}),
the authors have presented a mechanism to stabilize all the radions
by extending the Goldberger-Wise (GW) mechanism for the two 3-brane case
\cite{Goldberger:1999uk} and provided a solid ground to build 5D warped extra dimension models with multiple 3-branes.

\begin{figure}
\centering
\hspace{1.5cm}
   \includegraphics[width=0.6\linewidth]{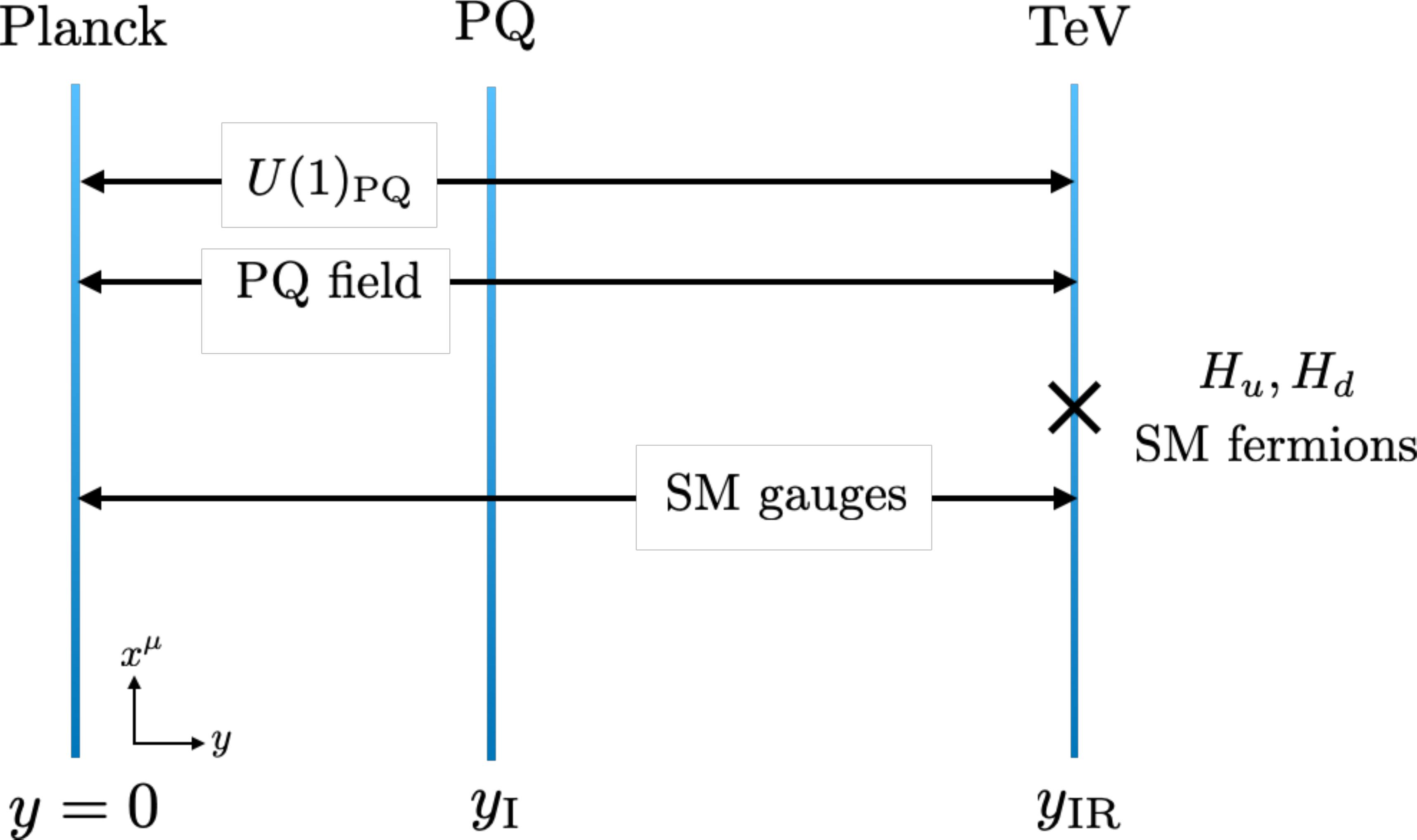}
   \vspace{0.2cm}
   \caption{A schematic picture of the setup.}
\label{fig:setup}
\end{figure}

In this paper, we consider a warped extra dimension model with three 3-branes
to explore a new possibility that provides a mechanism to address the electroweak naturalness problem
and an axion solution to the strong CP problem at the same time.
A schematic picture of our setup is summarized in Figure~\ref{fig:setup}.
The electroweak naturalness problem is addressed in the similar way as the RS model with two 3-branes.
We identify the typical scales of the UV and IR branes placed at orbifold fixed points in the extra dimension
as the Planck and TeV scales, and Higgs fields live on the TeV brane.
To solve the strong CP problem, we assume that the $U(1)_{\rm PQ}$ symmetry is realized as a gauge symmetry in the bulk of the extra dimension
(see refs.~\cite{Dienes:1999gw,Choi:2003wr,Flacke:2006ad,Cox:2019rro,Bonnefoy:2020llz,Yamada:2021uze}
for axion models introducing extra dimensions).
We introduce a 5D scalar field whose potential at the intermediate brane drives a spontaneous breaking of the $U(1)_{\rm PQ}$ symmetry.
Then, the $U(1)_{\rm PQ}$ breaking scale is given by the typical scale of the intermediate brane
and naturally small compared to the Planck scale due to a warp factor.
The axion profile is significantly suppressed around the UV brane so that the axion is protected from
gravitational violations of the $U(1)_{\rm PQ}$ symmetry on the UV brane and the quality problem is addressed
\cite{Cox:2019rro,Bonnefoy:2020llz}
(see ref.~\cite{Nakai:2021nyf} for a 4D realization with supersymmetry). 
We here employ the DFSZ axion model~\cite{Zhitnitsky:1980tq,Dine:1981rt} 
to introduce the axion-gluon coupling leading to the axion potential cancelling $\bar{\theta}$ at its minimum
(it is also possible to consider the KSVZ model~\cite{Kim:1979if,Shifman:1979if}).
Two doublet Higgs fields and SM fermions localized on the TeV brane are charged under the $U(1)_{\rm PQ}$ symmetry.
Cancellation of the $U(1)_{\rm PQ}$ gauge anomaly is achieved by the 5D Chern-Simons term.
Our model predicts Kaluza-Klein (KK) excitations of the QCD axion near the TeV scale
and relatively light extra Higgs bosons.

The rest of the paper is organized as follows.
In section~\ref{sec:set_up}, we present our warped extra dimension model with three 3-branes.
A $U(1)_{\rm PQ}$ breaking scalar field is introduced in the whole bulk region of a warped extra dimension.
We find the background profile of that scalar field by solving its equation of motion.
At the UV brane, the $U(1)_{\rm PQ}$ gauge symmetry is broken and $U(1)_{\rm PQ}$-violating interactions are expected.
In section~\ref{sec:axion}, we calculate the axion profile and mass generated from the explicit $U(1)_{\rm PQ}$ breaking
and show that the axion profile localized toward the intermediate brane can sequester the mass.
Section~\ref{sec:axion_gluon_coupling} discusses cancellation of the $U(1)_{\rm PQ}$ gauge anomaly
and shows the axion-gluon coupling in the low-energy effective theory leading to the desired axion potential.
We will see that the axion quality problem is solved. 
In section~\ref{KK_modes}, we explore predictions of the model.
Section~\ref{discussions} is devoted to conclusions and discussions.

\section{The model}
\label{sec:set_up}

We consider a warped extra dimension model with three 3-branes.
The $U(1)_{\rm PQ}$ gauge field and a $U(1)_{\rm PQ}$ breaking scalar field are introduced
in the whole bulk region of the extra dimension.
We solve the equation of motion for the scalar field and find the background profile.

\subsection{Three 3-branes}
\label{three_branes}

Our spacetime geometry is given by $\mathbb{R}^4\times S_1/\mathbb{Z}_2$,
and the metric is
\begin{align}
ds^2= g_{MN} dx^M dx^N = e^{-2\sigma(y)} \eta_{\mu\nu} dx^\mu dx^\nu-dy^2\ ,
\label{metric}
\end{align}
where $M=(\mu, y)$ and $x^{\mu \, (= 0,1,2,3)}$
denote the coordinates of the ordinary $4$-dimensional spacetime with the flat metric $\eta_{\mu\nu}$.
The coordinate of the $S_1/\mathbb{Z}_2$ orbifold is given by $y \in [y_{\rm UV} (=0), y_{\rm IR}]$
whose boundaries correspond to the fixed points of the $S_1/\mathbb{Z}_2$.
We introduce three 3-branes extending over $\mathbb{R}^4$ at $y=0, y_{\rm I}, y_{\rm IR}$
and call them the UV, intermediate and IR branes, respectively.
Then, the whole region of the extra dimension is divided into two subregions,
$0<y<y_{\rm I}$ (subregion 1) and $y_{\rm I}<y<y_{\rm IR}$ (subregion 2). 
A warp factor $\sigma (y)$ is given by
\begin{align}
\sigma (y)=
\left\{
\begin{array}{c}
k_1 y \quad \! \qquad \qquad \qquad \text{(subregion 1)}\\
k_2 y+(k_1-k_2)y_{\rm I} \quad \text{(subregion 2)}
\end{array}
\right.\ .
\end{align}
Here, $k_{1,2}$ are positive constants with a mass dimension.
We require them to satisfy $k_2 > k_1$ where radions are stabilized via the GW mechanism.
See ref.~\cite{Lee:2021wau} for a detailed discussion about the radion stabilization.

We now introduce a $U(1)_{\rm PQ}$ gauge field $A_M$
and a complex scalar field (the PQ field) $\Phi$ with $U(1)_{\rm PQ}$ charge $+1$
living in the whole 5D spacetime. 
The action of this system is given by
\begin{align}
S=\,\, &2\int^{y_{\rm IR}}_{0}d^5 x\sqrt{g}\left[-\frac{1}{4g_5^2}F^{MN}F_{MN}+\frac{1}{2g_5^2\xi}\left(g^{\mu\nu}\partial_\mu A_\nu+e^{2\sigma}\xi \partial_5(e^{-2\sigma} A_5)-\xi g_5^2|\Phi|^2 a \right)^2\right] \nonumber \\[1ex]
&+2\int^{y_{\rm I}}_{0}d^5 x\sqrt{g}\left[\frac{1}{2}|\mathcal{D}_M\Phi|^2-\frac{1}{2}m^2_{1}|\Phi|^2 \right]
+2\int^{y_{\rm IR}}_{y_{\rm I}}d^5 x\sqrt{g}\left[\frac{1}{2}|\mathcal{D}_M\Phi|^2-\frac{1}{2}m^2_{2}|\Phi|^2 \right]
\nonumber \\[1ex]
&-\sum_{{\rm i}={\rm UV,I,IR}}\int d^4 x \sqrt{g^{\rm in}_{\rm i}}\, U(\Phi)_{\rm i}\ .
\label{action}
\end{align}
Here, $g$ is the determinant of the metric \eqref{metric}, $F_{MN}=\partial_M A_N-\partial_N A_M$ is the field strength of the $U(1)_{\rm PQ}$,
$g_5$ is the 5D gauge coupling, $\xi$ is a gauge fixing parameter and
$a$ is a real scalar field defined by the decomposition of $\Phi \equiv(\langle \Phi\rangle +\eta)\,e^{ia}$
where $\langle \Phi\rangle$ denotes a vacuum expectation value (VEV) of $\Phi$ and $\eta$ is a real scalar field.
A gauge choice is determined by a value of $\xi$, $i.e.$ the $R_\xi$ gauge,
and mixings between the vector mode $A_\mu$, the scalar mode $A_5$ and $a$ are erased by the second term in the first line.
For the second line, the covariant derivative is defined as $\mathcal{D}_M\equiv \partial_M-i A_M$
and $m_{1,2}$ are real parameters with a mass dimension. 
In the last line,  $U(\Phi)_{\rm i}$ denotes a brane-localized potential of the PQ field
and $g^{\rm in}_{\rm i}\equiv e^{-8\sigma}|_{y=y_{\rm i}}$ is the determinant of the induced metric at $y=y_{\rm i}$.

To realize the QCD axion, the $U(1)_{\rm PQ}$ gauge field $A_\mu$ satisfies the Dirichlet boundary condition at the UV brane
where the $U(1)_{\rm PQ}$ symmetry is explicitly broken, while
the $U(1)_{\rm PQ}$ gauge symmetry is only spontaneously broken in the other regions of the extra dimension. 
Then, we assume the following brane-localized potentials at the UV and intermediate branes:
\begin{align}
&U(\Phi)_{\rm UV}= \left(-l_{\rm UV}k_1^{5/2}\Phi+c.c. \right)+b_{\rm UV}k_1 |\Phi|^2 \ , \label{UUV} \\[1ex]
&U(\Phi)_{\rm I}=\frac{\lambda_{\rm I}}{k_1^2} \left(|\Phi|^2-k_1^3v_{\rm I}^2 \right)^2\ , \label{UI}
\end{align}
where $l_{\rm UV}, b_{\rm UV}, \lambda_{\rm I}, v_{\rm I}$ are dimensionless constants. 
The first term in $U(\Phi)_{\rm UV}$ is the most dangerous term that we can write to break the $U(1)_{\rm PQ}$ explicitly.
The second mass term preserves the $U(1)_{\rm PQ}$ symmetry.
Both of \eqref{UUV} and \eqref{UI} induce a nonzero VEV for the PQ field.

While the SM gauge fields propagate in the whole bulk of the extra dimension,
we assume a simple setup that two $SU(2)_L$ doublet Higgs fields $H_u, H_d$ and the SM fermions live only on the IR (TeV) brane.
The Higgs fields couple to the PQ field on the IR brane,
\begin{align}
\label{eq:Phi2HuHd}
S \supset -\int d^4 x\sqrt{g^{\rm in}_{\rm IR}}\, \frac{\kappa_{ud}}{M_5}\,\Phi^2H_uH_d\ ,
\end{align}
where $\kappa_{ud}$ is a dimensionless constant and $M_5$ is the $5$D Planck mass.
This setup is similar to the DFSZ axion model~\cite{Zhitnitsky:1980tq,Dine:1981rt}
while the Higgs fields and the SM fermions are understood as composite states in the dual 4D CFT picture of the current extra dimension model.
Through the coupling \eqref{eq:Phi2HuHd}, the Higgs fields and the SM fermions are charged under the $U(1)_{\rm PQ}$.
We will discuss the cancellation of the $U(1)_{\rm PQ}$ gauge anomaly in a later section.

\subsection{The background profile}
\label{sec:background}

Let us compute the background solution of the PQ field $\langle \Phi \rangle$.
Our approach is similar to that of the model with two 3-branes~\cite{Cox:2019rro}.
We neglect the backreaction of the PQ field to the metric
by assuming that the energy density of the PQ field is much smaller than that obtained via the background geometry.
The bulk equation of motion for the subregion $p \, (=1,2)$ is given by
\begin{align}
&-\partial_5(e^{-4\sigma}\partial_5\langle\Phi\rangle)+m_{p}^2\, e^{-4\sigma}\langle\Phi\rangle=0\ .
\end{align}
Here, $\partial_5$ denotes the derivative with respect to $y$. Note that the PQ field $\Phi$ has different bulk masses $m_1, m_2$
for two subregions, which is important to obtain a sizable coupling of $\Phi^2 H_uH_d$ at the IR brane as we will see later.
The bulk solutions take the following form:
\begin{align}
\label{eq:Phi}
\begin{split}
&\langle\Phi_1\rangle=k_1^{3/2}\left[c_a\, e^{k_1y(4-\Delta_1)}+c_b\, e^{k_1y \Delta_1}\right] \quad (\text{subregion~$1$})\ ,\\[1ex]
&\langle\Phi_2\rangle=k_2^{3/2}\left[c_c\, e^{k_2y(4-\Delta_2)}+c_d\, e^{k_2y \Delta_2}\right] \quad (\text{subregion~$2$})\ ,
\end{split}
\end{align}
where $\langle\Phi_p\rangle$ denotes the background solution for $\Phi$ in the subregion $p$, $c_{a,b,c,d}$ are dimensionless constants,
and $\Delta_{1,2}$ are defined as $m_{1,2}^2\equiv\Delta_{1,2} \left(\Delta_{1,2}-4 \right)k_{1,2}^2$, respectively.
Requiring boundary terms in the variation of the action \eqref{action} to vanish,
the boundary conditions at the UV, intermediate and IR branes are
\begin{align}
&\partial_5\langle\Phi_1\rangle+l_{\rm UV}k_1^{5/2}-b_{\rm UV}k_1\langle\Phi_1\rangle |_{y=0}=0\ , \label{boundaryUV} \\
&[\partial_5\langle\Phi\rangle]|_{y=y_{\rm I}}-2\frac{\lambda_{\rm I}}{k_1^2} \left( |\langle\Phi_1\rangle|^2-k_1^3v_{\rm I}^2 \right) \langle\Phi_1\rangle |_{y=y_{\rm I}}=0\ ,
\label{boundaryI} \\
&[\langle\Phi\rangle]|_{y=y_{\rm I}}=0\ , \label{continuity} \\[1ex]
&\partial_5\langle\Phi_2\rangle|_{y=y_{\rm IR}}=0\ , \label{boundaryIR}
\end{align}
where $[X]|_{y=y_{\rm i}}\equiv \lim_{\epsilon\to 0+} X(x,y_{\rm i}+\epsilon)-X(x,y_{\rm i}-\epsilon)$
is defined for a function $X(x,y)$.
The third condition is the continuity of $\Phi$ at the intermediate brane.

\begin{figure}
\centering
   \includegraphics[width=0.46\linewidth]{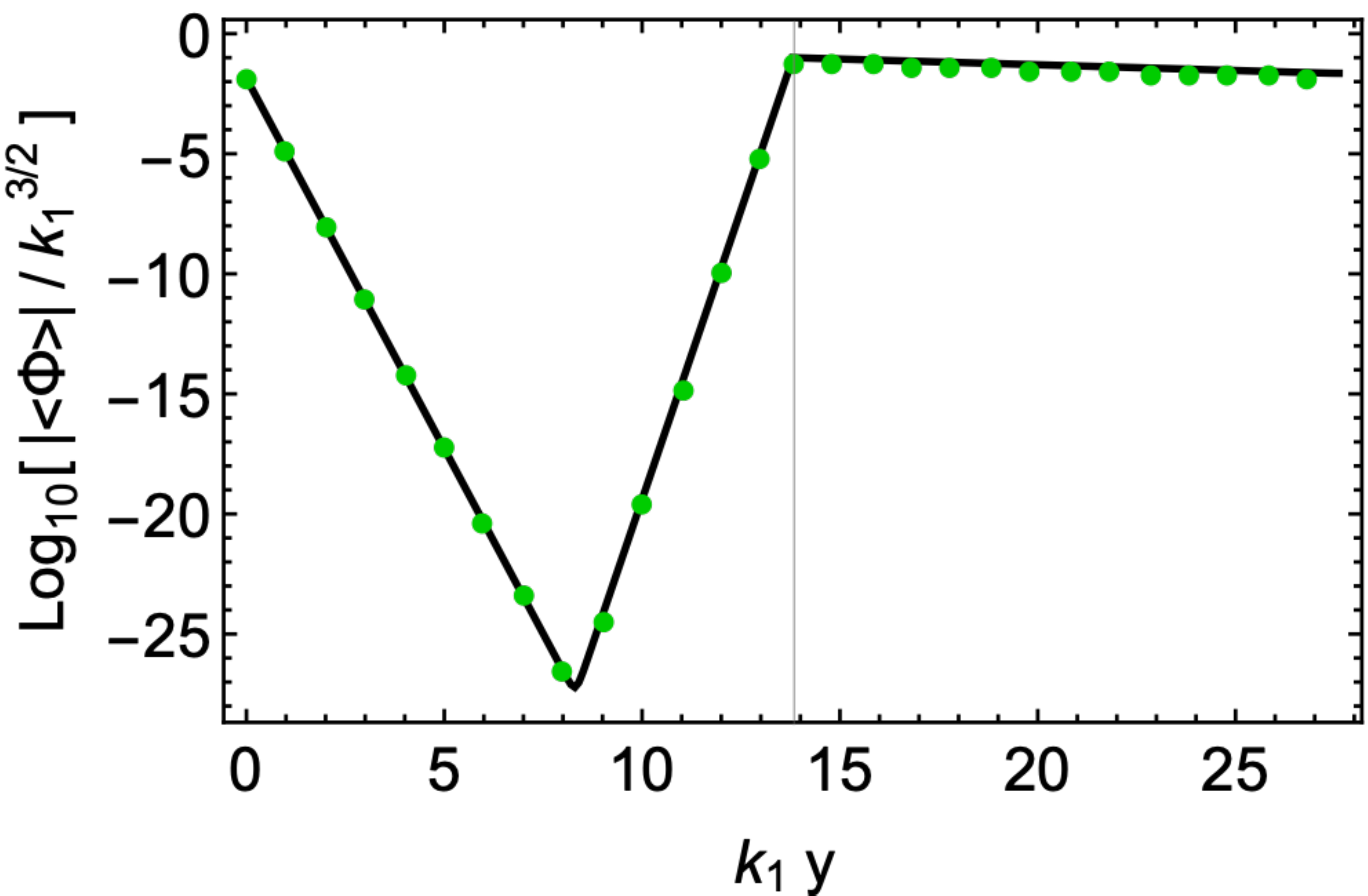}
   \hspace{0.8cm}
   \includegraphics[width=0.46\linewidth]{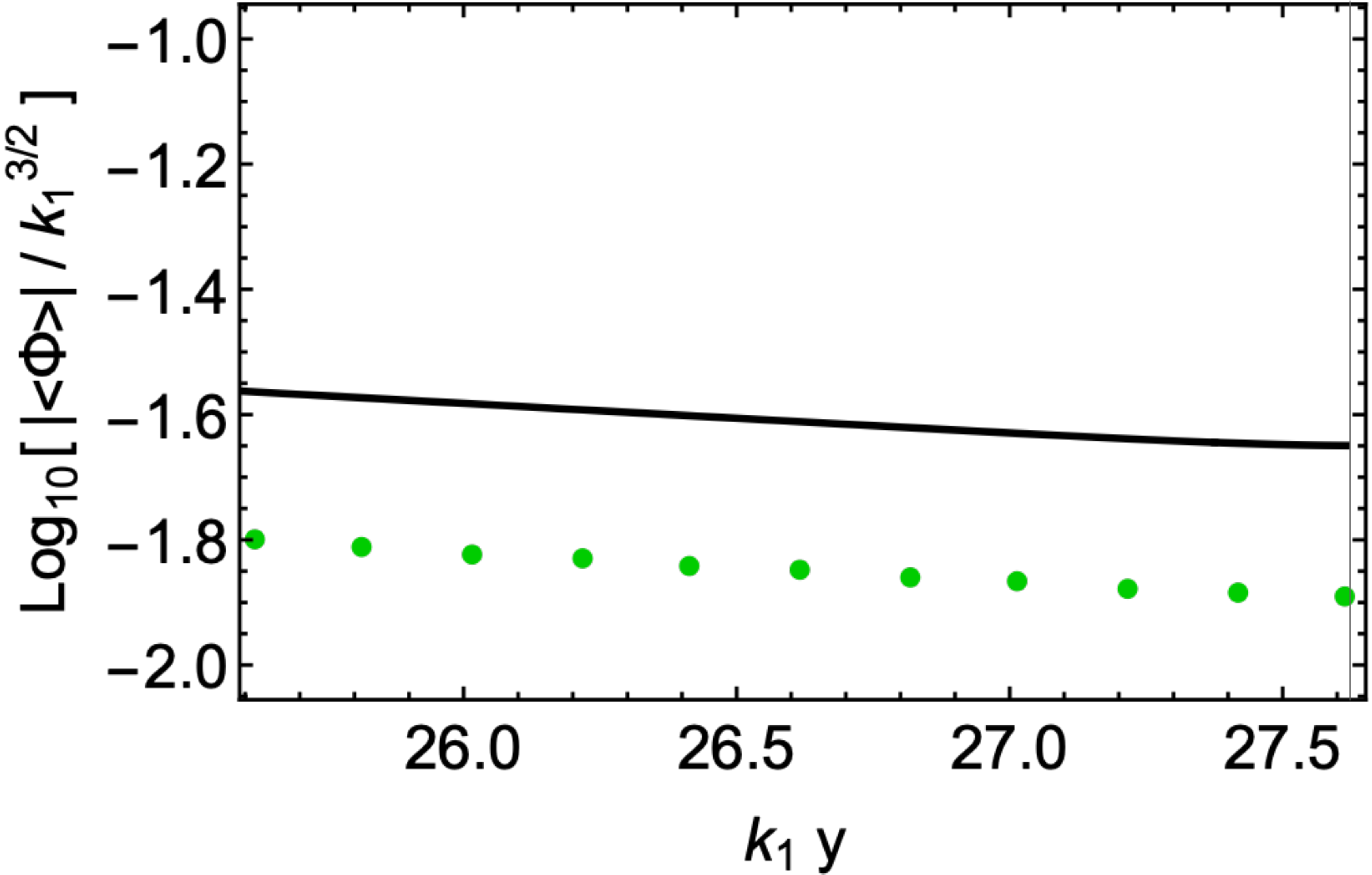}
   \caption{The normalized background profile of $|\langle\Phi\rangle|/k_1^{3/2}$
   for the whole region of the extra dimension (left) and the focused region close to the IR brane (right).
   The green dotted line denotes a numerical solution while
   the solid line is obtained by the approximated analytic formula.
   We here take $k_2/k_1=1.1, y_{\rm I}=\ln(10^6)/k_1, y_{\rm IR}=\ln(10^{13.2})/k_2, \Delta_1=11.1, \Delta_2=4.1, l_{\rm UV}=0.1, \lambda_{\rm I}=3.1, b_{\rm UV}=0.2$, and $\sigma_0=0.1$.
   The vertical thin line in the middle of the left panel gives the position of the intermediate brane.}
\label{fig:background}
\end{figure}

The background profile can be obtained by solving the equations numerically,
and the result for a reference parameter set is shown in Fig.~\ref{fig:background} (green dotted line).
Let us also derive the approximate analytic formula.
The boundary conditions \eqref{boundaryUV}, \eqref{boundaryIR} lead to
\begin{align}
\label{eq:cacc}
c_a=\frac{l_{\rm UV}+c_b(-b_{\rm UV}+\Delta_1)}{-4+b_{\rm UV}+\Delta_1}\ ,
\qquad c_c=\frac{c_d\,e^{2k_2y_{\rm IR}(\Delta_2-2)}\Delta_2}{\Delta_2-4}\ .
\end{align}
By substituting these constants into the bulk solutions \eqref{eq:Phi}
and using the continuity condition \eqref{continuity},
we obtain
\begin{align}
c_d = \,\, &\left(\Delta_2-4 \right) k_1^{3/2} e^{-(\Delta_1-4) k_1 y_{\rm I}} \nonumber \\
&\times \frac{ c_b \left\{b_{\rm UV} \left(e^{2 (\Delta_1-2) k_1
   y_{\rm I}}-1\right)+\Delta_1+(\Delta_1-4) e^{2 (\Delta_1-2)
   k_1 y_{\rm I}}\right\}+l_{\rm UV}}{k_2^{3/2} \left(b_{\rm UV}+\Delta_1-4 \right) \left\{ (\Delta_2-4) e^{\Delta_2 k_2
   y_{\rm I}}+\Delta_2 e^{k_2 (2 (\Delta_2-2)
   y_{\rm IR}-(\Delta_2-4) y_{\rm I})}\right\}}\ .
   \label{cd}
\end{align}
We are interested in the parameter space with $\Delta_1\gtrsim 10$, $e^{-k_1y_{\rm I}}\ll 1$ and
$ l_{\rm UV},b_{\rm UV},v_{\rm I},\lambda_{\rm I}\lesssim\mathcal{O}(1)$
where Eq.~\eqref{eq:cacc} is described by a simpler approximated form.
In addition, using Eq.~\eqref{eq:cacc} and the boundary condition \eqref{boundaryI}, we find
\begin{align}
\label{eq:approx_cd}
&c_d
   \approx \frac{c_b \left(\Delta_2-4 \right) k_1^{3/2} e^{\Delta_1 k_1
   y_{\rm I}+(\Delta_2-4) k_2 y_{\rm I}-2 (\Delta_2-2) k_2
   y_{\rm IR}}}{\Delta_2 k_2^{3/2}}\ , \\[2ex]
   &c_b \approx
\pm e^{-k_1y_{\rm I}\Delta_1} \sigma_0 \ , \quad
\sigma_0 \equiv \sqrt{\frac{k_2(4-\Delta_2)+k_1(2v_{\rm I}^2\lambda_{\rm I}-\Delta_1)}{2k_1\lambda_{\rm I}}} \ .
\end{align}
Note that there remains a freedom to choose the overall sign for the constant $c_b$, which originates from a $\mathbb{Z}_2$ symmetry,
$\Phi\to -\Phi$, when an explicit $U(1)_{\rm PQ}$ breaking term is ignored, $i.e.$ $l_{\rm UV}=0$.
We use the positive sign in the following discussion.
The approximate analytic formulas for $c_{a,b,c,d}$ are then summarized as
\begin{align}
\label{eq:c_abcd}
\begin{split}
&c_a \approx \frac{l_{\rm UV}}{-4+b_{\rm UV}+\Delta_1}\ , \qquad 
c_b \approx e^{-k_1y_{\rm I}\Delta_1}\sigma_0\ ,\\[1ex]
&c_c \approx \left( \frac{k_1}{k_2} \right)^{3/2}\sigma_0\,e^{k_2y_{\rm I}(\Delta_2-4)}\ , \qquad 
c_d \approx \left( \frac{k_1}{k_2} \right)^{3/2} \frac{e^{k_2y_{\rm I}(\Delta_2-4)-2k_2y_{\rm IR}(\Delta_2-2)} \left(\Delta_2-4 \right)\sigma_0}{\Delta_2}\ .
\end{split}
\end{align}
Fig.~\ref{fig:background} also shows the approximate analytic expression for the background profile (solid black line),
which agrees well to the numerical solution.
In our analysis, we focus on $\Delta_2\approx 4$ because
a larger $\langle\Phi\rangle$ at the IR brane is obtained for a smaller $\Delta_2 < 4$.
To neglect the backreaction of the PQ field to the metric
requires $|\langle\Phi\rangle'|^2+m_2^2|\langle\Phi\rangle|^2\ll 6k_2^2 M_5^3$ which leads to $\Delta_2 > 4-\ln(6M_5^3/\sigma_0^2k_1^3)/(2k_2(y_{\rm IR}-y_{\rm I}))$
($k_1\approx k_2$ is assumed for simplicity).
Besides, the Breitenlohner-Freedman bound $\Delta_{2}\geq 2$~\cite{Breitenlohner:1982bm} must be satisfied.
On the other hand, we get a smaller $\langle\Phi\rangle$ at the IR brane for a larger $\Delta_2 > 4$. 
This suppression of $\langle\Phi(y_{\rm IR})\rangle$ makes extra Higgs bosons in the two doublet Higgs fields dangerously light
as we will discuss in Sec.~\ref{sec:pheno}.
If we consider the typical mass scale of the IR brane less than $\mathcal{O}(10)$\,TeV, $\Delta_2\gg 4$ is 
disfavored 
due to $e.g.$ collider constraints.
These restrictions are satisfied for $\Delta_2\approx 4$
where $\langle \Phi_2 \rangle$ has an almost flat profile.

\section{The axion mass and profile}
\label{sec:axion}

Having obtained the background profile of the PQ field $\langle\Phi\rangle$,
we now explore the $U(1)_{\rm PQ}$ breaking mass and profile of the lightest mode of the axion $a$ and
the mass and profile of $A_5$ by solving their equations of motion.
The action \eqref{action} gives the following bulk equations,
\begin{align}
\label{eq:A5}
&\Box A_5 - g_5^2|\langle \Phi\rangle|^2e^{-2\sigma} \left(\partial_5a-A_5 \right)
+\xi \, \partial_5 \left\{ \partial_5(e^{-2\sigma}A_5)-g_5^2e^{-2\sigma}|\langle \Phi\rangle|^2a \right\}=0\ ,\\[1ex]
\label{eq:a}
&|\langle \Phi\rangle|^2\Box a- e^{2\sigma} \partial_5 \left\{e^{-4\sigma}|\langle \Phi\rangle|^2 \left(\partial_5a-A_5 \right) \right\}
+\xi |\langle \Phi\rangle|^2 \left\{\partial_5(e^{-2\sigma}A_5)-g_5^2|\langle \Phi\rangle|^2e^{-2\sigma}a \right\}=0\ .
\end{align}
The UV and IR boundary terms in the variation of the action \eqref{action} must vanish,
which leads to the boundary conditions at the UV and IR branes,
\begin{align}
&\delta a \left(2|\langle \Phi\rangle|^2 \left(\partial_5 a-A_5 \right)\mp \frac{\delta U_{\rm i}}{\delta a}\right)  \biggr|_{y=y_{\rm UV},y_{\rm IR}}=0\ ,
\label{deltaa} \\[1ex]
&\eta^{\mu\nu}\delta A_\mu \left(\partial_5A_\nu-\partial_\nu A_5 \right) \Bigr|_{y=y_{\rm UV},y_{\rm IR}}=0\ , \label{deltaAmu} \\[1ex]
&\delta A_5 \left\{e^{2\sigma}\eta^{\mu\nu}\partial_\mu A_\nu+e^{2\sigma}\xi \left( \partial_5(e^{-2\sigma}A_5)
-g_5^2e^{-2\sigma}|\langle \Phi\rangle|^2a \right) \right\}\Bigr|_{y=y_{\rm UV},y_{\rm IR}}=0\ , \label{deltaA5}
\end{align}
where $\delta a, \delta A_\mu, \delta A_5$ denote variations of $a, A_\mu, A_5$, respectively. 
While the gauge field $A_\mu$ is not included in the bulk equations of $A_5$ and $a$,
it appears in the boundary conditions \eqref{deltaA5} derived from the variation of $A_5$.
Then, we take account of the boundary conditions \eqref{deltaAmu} obtained through the variation of $A_\mu$.
Similarly, the boundary conditions at the intermediate brane are
\begin{align}
&\delta a [-2|\langle \Phi\rangle|^2 \left(\partial_5 a-A_5 \right) ] |_{y=y_{\rm I}}=\delta a \frac{\delta U_{\rm I}}{\delta a} \biggr|_{y=y_{\rm I}} \ ,
\label{intera} \\[1ex]
&\eta^{\mu\nu}\delta A_\mu [\partial_5 A_\nu-\partial_\nu A_5 ] |_{y=y_{\rm I}}=0\ , \label{interAmu} \\[2ex]
&\delta A_5 [\eta^{\mu\nu}\partial_\mu A_\nu
+\xi \{ \partial_5(e^{-2\sigma}A_5)-g_5^2e^{-2\sigma}|\langle \Phi\rangle|^2a \} ] |_{y=y_{\rm I}}=0\ . \label{interA5}
\end{align}
Here, we have used $\lim_{\epsilon\to 0+}\left[\delta X(x,y_{\rm i}+\epsilon)-\delta X(x,y_{\rm i}-\epsilon)\right] = 0$ for a function $X(x,y)$, 
and $\frac{\delta U_{\rm i}}{\delta a}$ denotes the derivative of the boundary potential $U_{\rm i}$ with respect to $a$,
\begin{align}
\frac{\delta U_{\rm UV}}{\delta a}=2l_{\rm UV}k_1^{5/2}\langle\Phi\rangle a \Bigr|_{y=y_{\rm UV}} \ ,
\qquad \frac{\delta U_{\rm I}}{\delta a}=\frac{\delta U_{\rm IR}}{\delta a}=0\ ,
\end{align}
where we have ignored the Higgs VEVs and only the explicit $U(1)_{\rm PQ}$ breaking term on the UV brane gives a non-trivial contribution.
Then, the conditions \eqref{deltaa}, \eqref{deltaAmu}, \eqref{deltaA5} can be reduced to the following boundary conditions at the UV brane,
\begin{align}
\label{aUVcondition}
&|\langle \Phi\rangle|^2 \left(\partial_5 a-A_5 \right)-l_{\rm UV}k_1^{5/2}\langle\Phi\rangle a \Bigr|_{y=y_{\rm UV}}=0\ ,\\[0.5ex]
\label{eq:bc_uv_3}
&A_\mu|_{y=y_{\rm UV}}=0\ , \\[1ex]
&\partial_5 \left(e^{-2\sigma}A_5 \right)-g_5^2e^{-2\sigma}|\langle \Phi\rangle|^2a \Bigr|_{y=y_{\rm UV}}=0\ . \label{A5_UV}
\end{align}
The condition \eqref{eq:bc_uv_3}, $i.e.$ the Dirichlet condition on $A_\mu$, is required to forbid a massless mode for $A_\mu$.
On the other hand, we impose the following boundary conditions at the IR brane,
\begin{align}
&\partial_5 a|_{y=y_{\rm IR}}=0\ , \label{aIRcondition} \\[1ex]
&\partial_5A_\mu|_{y=y_{\rm IR}}=0\ ,\\[1ex]
\label{eq:bc_IR_3}
&A_5|_{y=y_{\rm IR}}=0\ .
\end{align}
Finally, the boundary conditions at the intermediate brane are chosen to satisfy the conditions \eqref{intera}, \eqref{interAmu}, \eqref{interA5}:
\begin{align}
\label{continuitya}
&[a]|_{y=y_{\rm I}}=0\ , \qquad [\partial_5 a]|_{y=y_{\rm I}}=0\ ,\\[1ex]
\label{eq:bc_I_3}
&[A_\mu]|_{y=y_{\rm I}}=0\ , \qquad [\partial_5 A_\nu]|_{y=y_{\rm I}}=0\ ,\\[1ex]
&[A_5]|_{y=y_{\rm I}}=0\ , \qquad [\partial_5 (e^{-2\sigma}A_5 )]|_{y=y_{\rm I}}=0\ . \label{A5_intermediate}
\end{align}
Note that we do not impose the Dirichlet boundary conditions on $A_\mu$ at the intermediate and IR branes,
which enables to make the $U(1)_{\rm PQ}$ gauge symmetry only spontaneously broken in the whole space
except for the UV brane to solve the axion quality problem
as we will demonstrate later.
Furthermore, the Dirichlet boundary conditions on $A_5$ at the UV and intermediate branes
are not imposed because otherwise a trivial solution $a=A_5=0$ for the whole space is obtained.

Let us now solve the bulk equations \eqref{eq:A5}, \eqref{eq:a} with the boundary conditions.
We perform the KK decomposition for $a$,
\begin{align}
a(x,y)=\sum_n f^{(n)}_{a}(y)\,a^{(n)}(x)\ ,
\end{align}
where $\Box a^{(n)}(x)=-m_{a}^{(n)\,2}a^{(n)}(x)$. The eigenfunctions $a^{(n)}(x)$ are all orthogonal.
To solve the bulk equations, we expand them in terms of $g_5\sqrt{k_1}$ and assume $k_1\approx k_2$.
At the zeroth order of $g_5\sqrt{k_1}$, it is found that $a$ has a non-trivial solution while $A_5$ is trivial, $A_5=0$, for the whole space. 
In fact, Eq.~\eqref{eq:A5} is trivially satisfied at the zeroth order, and Eq.~\eqref{eq:a} is reduced to
  \begin{align}
  \label{eq:a_only}
&|\langle \Phi \rangle|^2m_{a}^2 \,f_a+ e^{2\sigma}\partial_5 \left\{e^{-4\sigma}|\langle \Phi \rangle|^2(\partial_5\, f_a) \right\}=0\ ,
\end{align}
for the mode $f_a(y)$ with mass eigenvalue $m_a$ (the label $n$ is omitted).
Using the background solutions \eqref{eq:Phi}, this equation is solved as
\begin{align}
\begin{split}
\label{eq:a1}
&f_{a, 1} =\frac{e^{k_1y(2-\Delta_1)}\left(J_{\Delta_1-2}\left(\frac{e^{k_1y}m_a}{k_1}\right)C_{a1}+Y_{\Delta_1-2}\left(\frac{e^{k_1y}m_a}{k_1}\right)C_{a2}\right)}{c_ae^{2k_1y(2-\Delta_1)}+c_b} \ , \\[2ex]
&f_{a, 2} =\frac{e^{k_2y(2-\Delta_2)}\left(J_{\Delta_2-2}\left(\frac{e^{k_2y+y_{\rm I}(k_1-k_2)}m_a}{k_2}\right)C_{a3}+Y_{\Delta_2-2}\left(\frac{e^{k_2y+y_{\rm I}(k_1-k_2)}m_a}{k_2}\right)C_{a4}\right)}{c_ce^{2k_2y(2-\Delta_2)}+c_d}\ ,
\end{split}
\end{align}
for the subregions $1, 2$.
Here, $J_n$ and $Y_n$ denote the $n$-th order Bessel functions of the first and second kind,
and $C_{a1,a2,a3,a4}$ are constants.

\begin{figure}
\centering
   \includegraphics[width=0.5\linewidth]{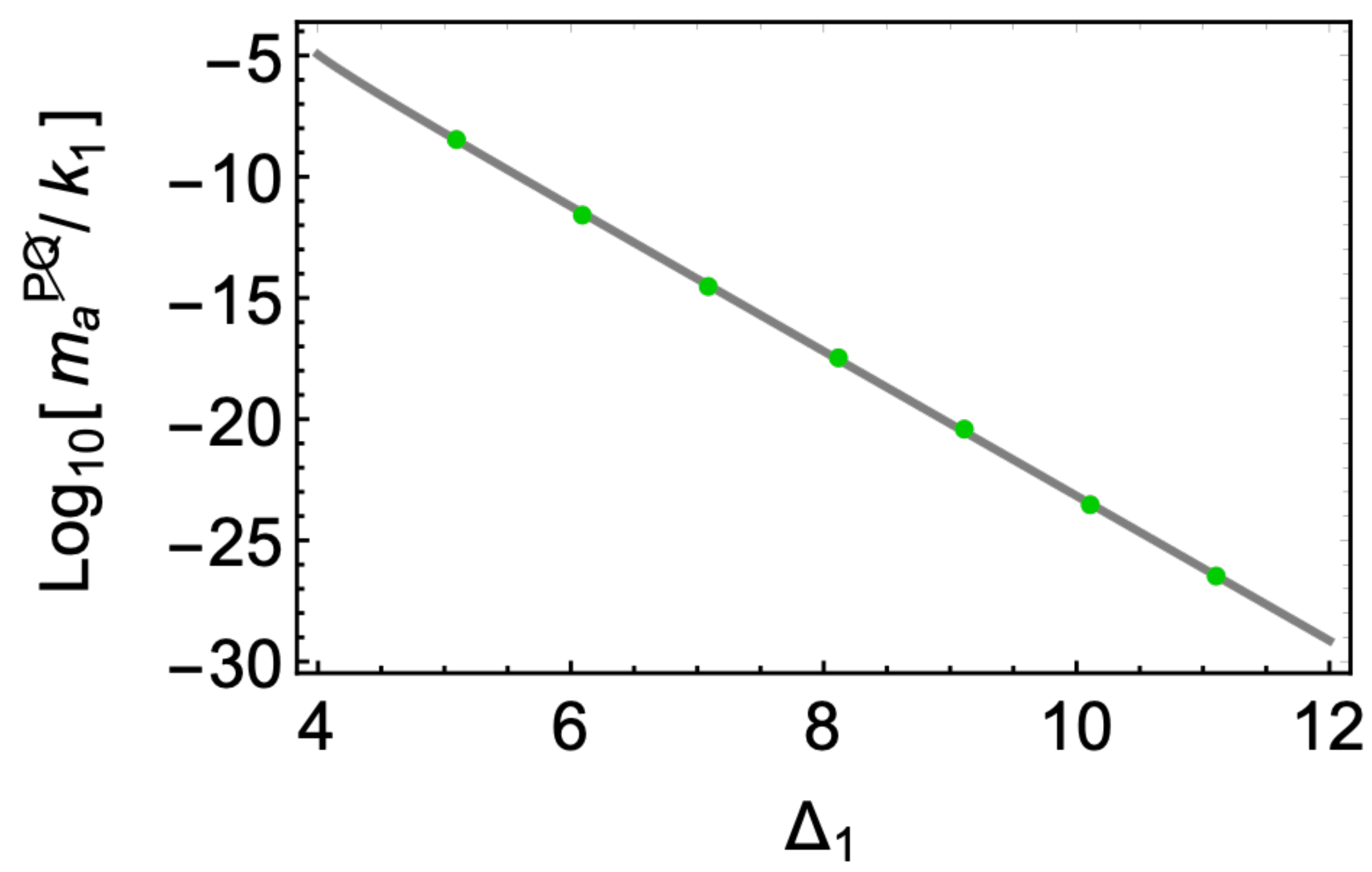}
   \caption{$m_a^{\cancel{\rm PQ}}/k_1$ as a function of $\Delta_1$.
  Here, we use the same parameter set as that of Fig.~\ref{fig:background}.
   The green dots and solid black line denote the results via the numerical computation of Eq.~\eqref{eq:a1} with Eq.~\eqref{eq:J_expand} and Eq.~\eqref{eq:c_abcd}
   and the analytical formula \eqref{eq:axion_mass_analytic}, respectively.
   }
\label{fig:ma_UV}
\end{figure}

We focus on the solution for the lightest mode here and discuss the solutions for the KK modes in Sec.~\ref{KK_modes}.
To solve the quality problem, we are interested in the parameter space
to achieve $m_a\ll k_1e^{-k_1y_{\rm I}}, k_2e^{-k_2y_{\rm IR}-y_{\rm I}(k_1-k_2)}$.
In this case, we can approximate the Bessel functions in Eq.~\eqref{eq:a1} as
\begin{align}
\begin{split}
\label{eq:J_expand}
&J_{\Delta-2}(x) \approx \frac{2^{-\Delta -3} x^{\Delta -2} \left\{ 32 \left(\Delta -1 \right) \Delta +x^4-8 \Delta 
   x^2\right\}}{\Gamma (\Delta +1)}\ ,\\[1ex]
&Y_{\Delta-2}(x) \approx -\frac{1}{\pi} \biggl[ 2^{\Delta -2} \Gamma (\Delta -2) x^{2-\Delta }  \\
&\qquad \qquad \qquad \quad +2^{-\Delta -3} \cos (\pi  \Delta )
   \Gamma (-\Delta ) \left\{32 \left(\Delta -1 \right) \Delta +x^4-8 \Delta  x^2\right\} x^{\Delta -2} \biggr] \ ,
\end{split}
\end{align}
where $\Delta\geq 4$ as mentioned before, $\Gamma (x)$ denotes the Gamma function
and higher order terms of the argument $x$ are neglected.
We use these expressions in Eq.~\eqref{eq:a1}.
The coefficients $C_{a1}, C_{a4}$ in Eq.~\eqref{eq:a1}
are determined by the boundary conditions at the UV and IR branes \eqref{aUVcondition}, \eqref{aIRcondition}.
Then, from the continuity of $a$ at the intermediate brane in Eq.~\eqref{continuitya},
the coefficient $C_{a3}$ is given in terms of $C_{a2}$ and $m_a$.
The remaining $C_{a2}$ is an overall factor in $f_a$.
The last boundary condition of the continuity of $\partial_5 a$ in Eq.~\eqref{continuitya} leads to the lightest axion mass. 
Using the approximated coefficients $c_{a,b,c,d}$ in Eq.~\eqref{eq:c_abcd} and assuming $\Delta_2=4$,
the equation to determine $m_a$ is given by
\begin{align}
0=\,\,&\frac{\Delta_1 l_{\rm UV}^2 
\mathcal{Z}^2_{1,0,0}
 \left\{32 \left(\Delta_1^2-3 \Delta_1+2\right) k_1^4-8 \left(\Delta_1-1 \right) k_1^2
   \widetilde m^2+\widetilde m^4\right\}}{b_{\rm UV}+\Delta_1-4} \nonumber \\[1ex]
  &+l_{\rm UV}
   \sigma_0 
   \mathcal{Z}_{-\Delta_1,0,0}
      \biggl[-\frac{16
   \Delta_1 \left(\Delta_1^2-3 \Delta_1+2\right) k_1^2
   \widetilde m^2 
  \mathcal{Z}^2_{1,0,0} 
   }{b_{\rm UV}+\Delta_1-4} \nonumber  \\[1ex]
   &+\frac{\Delta_1-2}{b_{\rm UV}+\Delta_1-4} \, 
  \mathcal{Z}^2_{-1,0,0}
   \Bigl\{32 \left(\Delta_1-1 \right) \Delta_1 k_1^4 \left( b_{\rm UV}+\Delta_1-4 \right) 
 \mathcal{Z}^4_{1,0,0} 
\nonumber  \\[1ex]
   &+8 \Delta_1 k_1^2 \widetilde m^2 \left(-b_{\rm UV}+\Delta_1+2 \right)
 \mathcal{Z}^2_{1,0,0} 
 +\widetilde m^4 \left(b_{\rm UV}-\Delta_1-4 \right) \Bigr\} \nonumber \\[1ex]
   &+\frac{2 \left(\Delta_1-2 \right)
   \Delta_1 \widetilde m^4
   \mathcal{Z}^2_{1,0,0} 
    }{b_{\rm UV}+\Delta_1-4}-8 \Delta_1 k_1^2 \widetilde m^2
  \mathcal{Z}^2_{\Delta_1-1,0,0} 
      +2 \widetilde m^4
  \mathcal{Z}^2_{\Delta_1-1,0,0} 
   \biggr] \nonumber \\[1ex]
   &+4
   \left(\Delta_1-2 \right) \sigma_0^2 \widetilde m^2 
   \mathcal{Z}^2_{-(\Delta_1+1),0,0} 
   \left\{4 \Delta_1 k_1^2 \left(
  \mathcal{Z}^2_{1,0,0}
   -
   \mathcal{Z}^2_{\Delta_1,0,0}
   \right)+\widetilde m^2 \left(
   \mathcal{Z}^2_{\Delta_1,0,0}
   -1\right)\right\} \ .
\end{align}
Here, we have defined $\mathcal{Z}_{a,b,c}\equiv e^{k_1y_{\rm I}a+k_2y_{\rm I}b+k_2y_{\rm IR}c}$
and $\widetilde m\equiv m_a\, e^{k_1y_{\rm I}}$.
Focusing on the parameter space with $\Delta_1\gg 4$, $e^{k_1y_{\rm I}}\ll e^{k_2y_{\rm IR}}$
and $m_a\ll k_2\,e^{-k_2y_{\rm IR}-(k_1-k_2)y_{\rm I}}$,  the above equation leads to
\begin{align}
\label{eq:axion_mass_analytic}
m^{\cancel{\rm PQ}}_a \approx 2 k_1 e^{-k_1y_{\rm I}(\Delta_1-2)/2}
\sqrt{\frac{ l_{\rm UV}}{\sigma_0} \frac{ 2-3\Delta_1+\Delta_1^2 }{b_{\rm UV}+\Delta_1 -4}}\ ,
\end{align}
where the superscript $\cancel{\rm PQ}$ is added for $m_a$ to denote the axion mass
originated from the explicit $U(1)_{\rm PQ}$  breaking term on the UV brane.
Fig.~\ref{fig:ma_UV} shows the numerical (green dots) and approximate analytic (black solid) results of $m_a^{\cancel{\rm PQ}}/k_1$
as a function of $\Delta_1$ for a representative parameter set.
The figure implies that the simple expression \eqref{eq:axion_mass_analytic} is valid even for $\Delta_2\lesssim 4.1$
while Eq.~\eqref{eq:axion_mass_analytic} was derived by assuming $\Delta_2=4$.
The $U(1)_{\rm PQ}$ breaking mass of the axion $m^{\cancel{\rm PQ}}_a$ is significantly suppressed for a larger $\Delta_1$.
In Sec.~\ref{sec:axion_gluon_coupling},
we will also discuss the axion mass $m_a^{\rm QCD}$ generated from non-perturbative QCD effects
and find a parameter space
where $m_a^{\cancel{\rm PQ}}$ is sufficiently suppressed compared to $m_a^{\rm QCD}$ and the axion quality problem is solved.

\begin{figure}
\centering
   \includegraphics[width=0.5\linewidth]{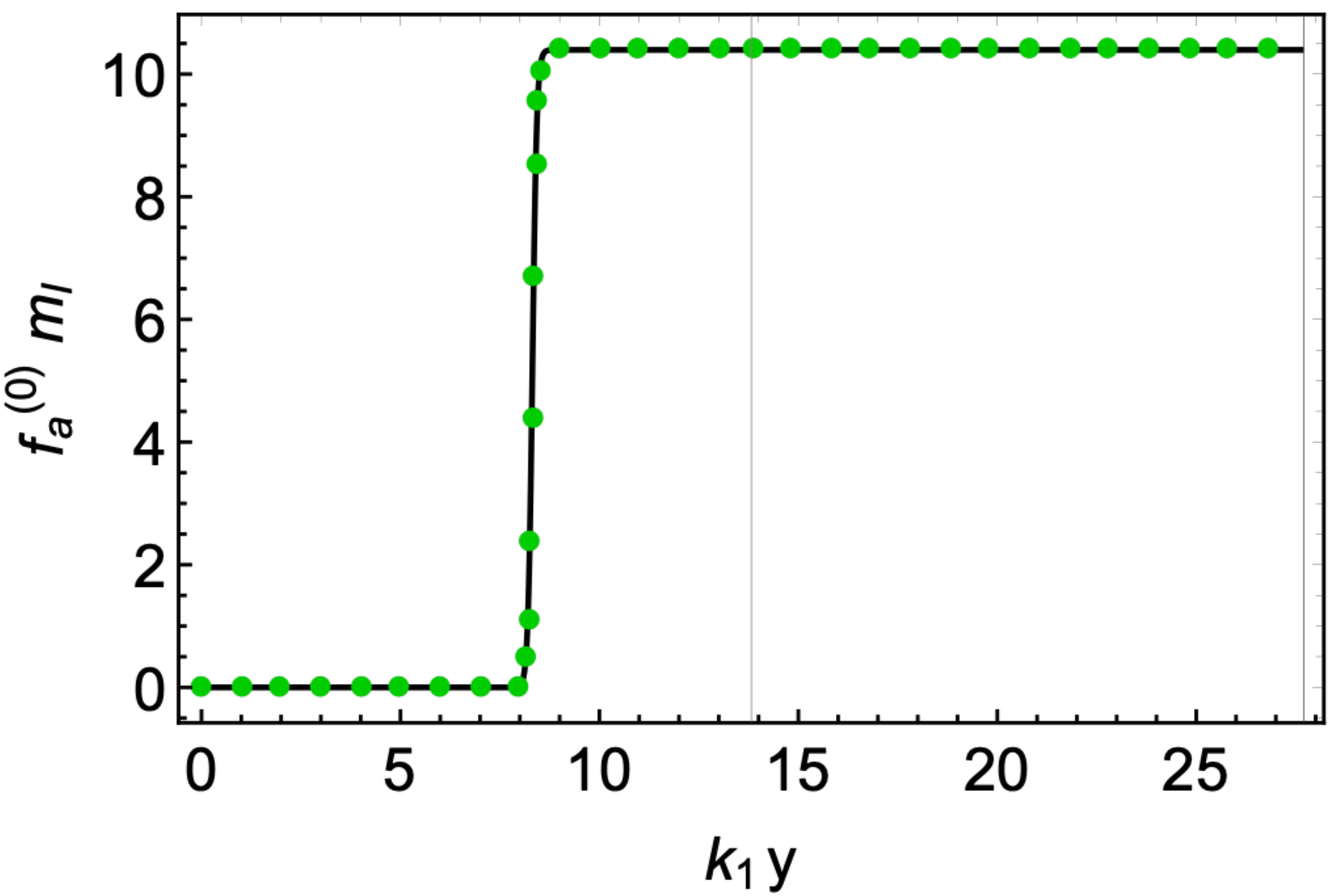}
   \caption{The numerical (green dots) and approximate analytic (black solid) results of the profile of the axion lightest mode, $f_a^{(0)}(y)$.
   We take 
   $m_{\rm I} \equiv k_1e^{-k_1y_{\rm I}}$
   and 
   the parameters 
   are the same as those used in Fig.~\ref{fig:background}.
   The left and right thin vertical lines denote the positions of the intermediate and IR branes, respectively.
   }
\label{fig:axion_profile}
\end{figure}

The overall factor $C_{a2}$ in $f_a$ is fixed by canonically normalizing the axion field $a$.
The axion kinetic terms are contained in
\begin{align}
\label{eq:axion_kin}
S\supset 2\int_{y_{\rm UV}}^{y_{\rm IR}}d^5 x\sqrt{g}\left(\frac{1}{2} \langle\Phi\rangle^2a\Box a+\frac{1}{2g_5^2}A_5\Box A_5\right) .
\end{align}
The contribution from the second term including $A_5$ is negligible
when we focus on the leading order of $g_5\sqrt{k_{1,2}}$.
Then, ignoring the second term, the condition of the canonical normalization is given by
\begin{align}
2\int_{y_{\rm UV}}^{y_{\rm IR}} dy\sqrt{g} \, \frac{1}{2}\langle\Phi\rangle^2 f_a^{(n)\,2}=\frac{1}{2} \ ,
\end{align}
which determines the constant $C_{a2}$.
Fig.~\ref{fig:axion_profile} shows the numerical (green dots) and approximate analytic (black solid) results
of the profile of the axion lightest mode, $f_a^{(0)}(y)$, for a reference parameter set.
The approximate analytic profiles are given by
\begin{align}
f_{a,1}&\approx \frac{\left(b_{\rm UV}+\Delta_1-4 \right) 
   e^{k_1 (2 (\Delta_1-2) y+y_{\rm I})}}{k_1 \sigma_0
   \left(b_{\rm UV}+\Delta_1-4 \right) e^{2 (\Delta_1-2) k_1 y}+k_1
   l_{\rm UV} e^{\Delta_1 k_1 y_{\rm I}}} \sqrt{\frac{\left(\Delta_1-1 \right) \left(\Delta_2-3 \right) k_2}{\left(\Delta_1-1 \right) k_1
   + \left(\Delta_2-3 \right) k_2}}
   \ , \nonumber \\[1ex]
f_{a,2}&\approx \frac{e^{k_1 y_{\rm I}}}{k_1 \sigma_0} \sqrt{\frac{\left(\Delta_1-1 \right) \left(\Delta_2-3 \right)
   k_2}{\left(\Delta_1-1 \right) k_1+ \left(\Delta_2-3 \right)
   k_2}}
 \ ,    
\label{axionprofileanalytic}
\end{align}
for the subregions $1,2$.
They agree well to the numerical result.
The figure shows that the profile is suppressed around the UV brane, experiences an exponential growth
and saturates before reaching the intermediate brane.
The almost flat profile is maintained until the IR brane.
This behavior can be easily read from the analytic expressions. 
The suppression of $f_a^{(0)}(y)$ around the UV brane indicates the suppression of $m_a^{\cancel{\rm PQ}}$
originated from the explicit $U(1)_{\rm PQ}$ breaking term on the UV brane.

The KK decomposition for the field $A_5$ is given by $A_5=\sum_n f_{A_5}^{(n)}(y)\,A_5^{(n)}(x)$.
At the first order of $g_5\sqrt{k_{1,2}}$, the bulk equation \eqref{eq:A5} then gives the solutions,
\begin{align}
&f_{A_{5,1}}=g_5\sqrt{k_1} e^{2 k_1y} \left\{D_1\,I_0\left(\frac{e^{
   k_1y} m_a}{k_1\sqrt{\xi }}\right)+2  D_2\, K_0\left(\frac{e^{ k_1y} m_a}{k_1\sqrt{\xi }}\right)\right\} \ , \nonumber \\[1ex]
&f_{A_{5,2}}=g_5 \sqrt{k_2} e^{2 k_2 y} \left\{D_3\, I_0\left(\frac{e^{
   k_2 y+ (k_1 - k_2) y_{\rm I}} m_a}{k_2 \sqrt{\xi
   }}\right)+2 D_4\, K_0\left(\frac{e^{k_2 y+ (k_1 -
   k_2) y_{\rm I}} m_a}{k_2 \sqrt{\xi }}\right)\right\} \ ,
\end{align}
for the subregions $1,2$.
Here, $I_0$ and $K_0$ are the modified Bessel functions of the first and second kind and $D_{1,2,3,4}$ are constants
determined by the boundary conditions \eqref{A5_UV}, \eqref{eq:bc_IR_3}, \eqref{A5_intermediate}.
However, they lead to $D_{1,2,3,4}=0$, and thus we find $f_{A_5}=\mathcal{O}(g_5^2k_{1,2})$.
The similar result has been also reported in the two 3-brane model~\cite{Cox:2019rro}.

\section{Axion quality}
\label{sec:axion_gluon_coupling}

Under the $U(1)_{\rm PQ}$ gauge transformation on the IR brane, $i.e.$ $ A_\mu \to A_\mu+\partial_\mu\alpha(x,y_{\rm IR})$,  $\Phi \to \Phi\, e^{i\alpha(x,y_{\rm IR})}$ and the corresponding transformations of the SM fermions and the two Higgs doublets,
the following term is generated due to the $U(1)_{\rm PQ}-SU(3)_c^2$ gauge anomaly,
\begin{align}
\label{gaugeanomaly}
\delta S\supset \frac{A_{\rm QCD}}{32\pi^2}\int d^4 x\, \alpha (x,y_{\rm IR}) \, G^a_{\mu\nu} \widetilde G^{a\mu\nu}\ ,
\end{align}
where $G^a_{MN}$ ($a=1 \sim 8$) is the field strength of the 5D color $SU(3)_c$ gauge field,
$G^a_{\mu\nu}$ are its 4D components and 
the dual is defined by $\widetilde G^{a\mu\nu} \equiv \frac{1}{2}\epsilon^{\mu\nu\rho\sigma}G^a_{\rho\sigma}$.
The coefficient of the term is calculated as $A_{\rm QCD}=6$.
This gauge anomaly term can be canceled by including the 5D Chern-Simons term in the bulk action,
\begin{align}
S \supset \frac{\kappa}{32\pi^2}\int_{y_{\rm UV}}^{y_{\rm IR}}d^5 x\,\epsilon^{MNPQR}A_M G_{NP}^aG^a_{QR}\ .
\end{align}
Here, $\kappa$ is a dimensionless constant and $\epsilon^{MNPQR}$ is the 5D Levi-Civita tensor density.
Under the 5D gauge transformation of $ A_M\to A_M+\partial_M\alpha(x,y)$, the Chern-Simons term generates the boundary terms,
\begin{align}
\label{CSanomaly}
\delta S \supset \frac{\kappa}{32\pi^2}\left[\int d^4x\,\alpha(x,y)\,G^a_{\mu\nu} \widetilde G^{a\mu\nu} \right]_{y_{\rm UV}}^{y_{\rm IR}}\ .
\end{align}
The gauge anomaly on the IR brane \eqref{gaugeanomaly} is cancelled by choosing $\kappa=-A_{\rm QCD}$.
Note that Eq.~\eqref{CSanomaly} also generates the gauge anomaly on the UV brane but it is irrelevant
because the $U(1)_{\rm PQ}$ symmetry is explicitly broken at the UV brane.
The other gauge anomalies, $e.g.$ $U(1)_Y-U(1)_{\rm PQ}^2$, $U(1)_{\rm PQ}^3$, $U(1)_{\rm PQ}-U(1)_{Y}^2$, on the IR brane
are canceled in a similar manner.

\begin{figure}
\centering
\includegraphics[width=0.4\linewidth]{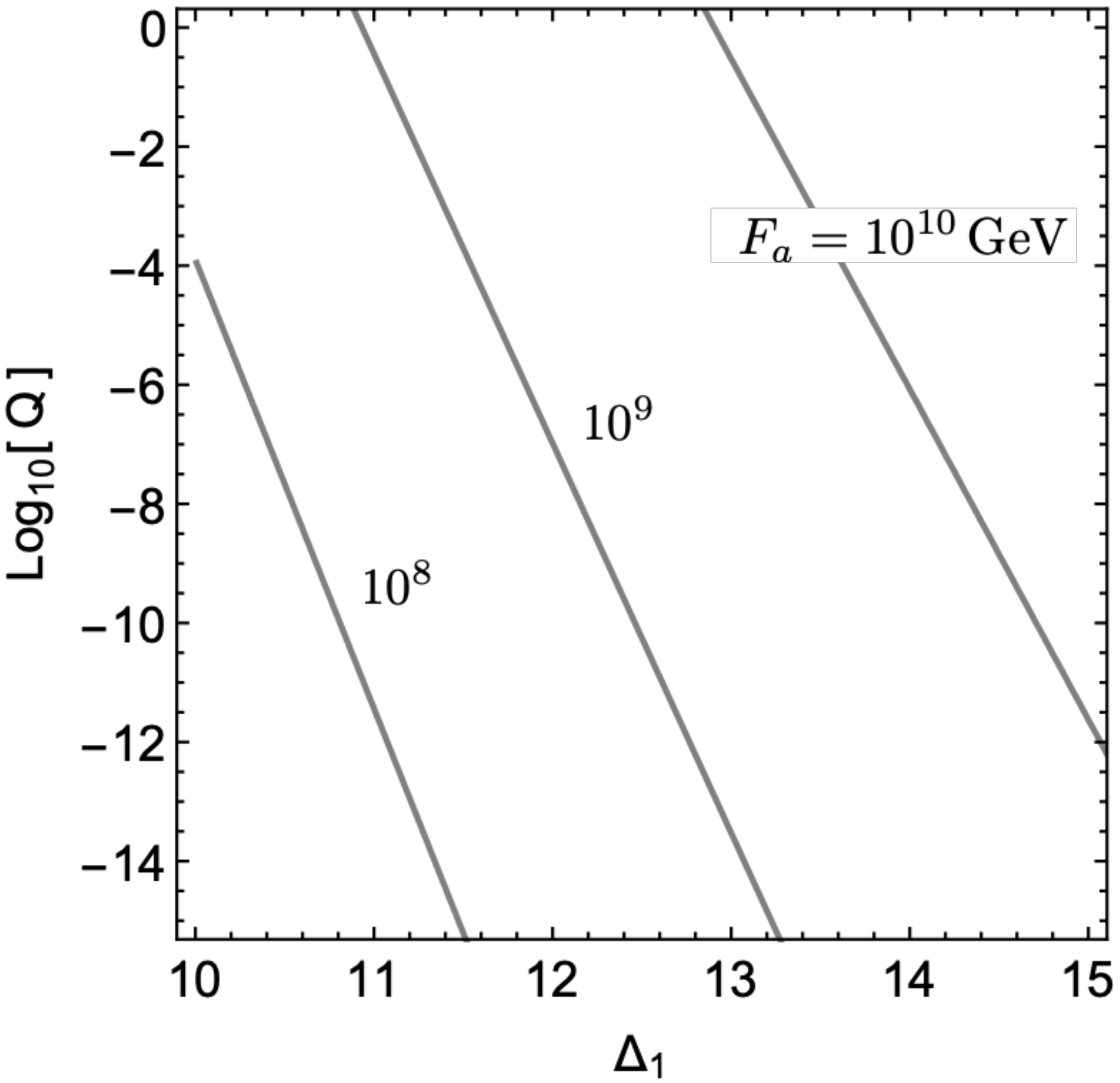}
   \caption{
   The contours of the quality factor $\mathcal{Q}$ as a function of $\Delta_1$
for $F_a = 10^{8,9,10} \, \rm GeV$.
    We take $k_1=0.1M_{\rm Pl}, k_2=1.1k_1, 
        \sigma_0=0.1, b_{\rm UV}=2.1, l_{\rm UV}=0.1$ and $\Delta_2=4$.
   }
\label{fig:quality}
\end{figure}

Due to the absence of the zero mode for the $U(1)_{\rm PQ}$ gauge field $A_M$,
the 4D effective theory at low energies after integrating out the extra dimension just reduces to the DFSZ axion model
that respects the approximate $U(1)_{\rm PQ}$ global symmetry.
The 4D effective action of the lightest mode of the axion $a^{(0)}(x)$ coupling to the 4D Higgs fields is derived from Eq.~\eqref{eq:Phi2HuHd} as
\begin{align}
S_{\rm eff} \supset-\int d^4x \, \frac{\kappa_{ud}e^{-2\sigma(y_{\rm IR})}\langle\Phi(y_{\rm IR})\rangle^2}{M_5} \,
e^{2if_a^{(0)}(y_{\rm IR})\, a^{(0)}(x)} H_uH_d\ .
\end{align}
Here, $\sigma(y_{\rm IR})$ is the warp factor at the IR brane, the Higgs fields $H_{u,d}$ are canonically normalized
and the axion profile $f_a^{(0)}(y_{\rm IR})=f_{a, 2} (y_{\rm IR})$ is given in Eq.~\eqref{axionprofileanalytic}.
As in the DFSZ model, we obtain the axion-gluon coupling leading to the axion potential to set $\bar{\theta}$ to zero
is obtained by redefining the SM fermions
to erase the axion dependence from the fermion mass terms,
\begin{align}
S_{\rm eff} \supset \int d^4 x\,\frac{1}{32\pi^2 F_a}a^{(0)}(x)G^{a(0)}_{\mu\nu} \widetilde G^{a(0)\mu\nu} \ ,
\qquad
\frac{1}{F_a} \equiv A_{\rm QCD}\,f^{(0)}_a(y_{\rm IR})\ ,
\end{align}
where $G^{a(0)}_{\mu\nu}$ denotes the field strength of the gluon zero mode.
Eq.~\eqref{axionprofileanalytic} indicates that the axion decay constant is given by the typical mass scale of the intermediate brane,
$F_a \sim k_1 e^{-k_1y_{\rm I}}$.
It is worth noting that the axion couples to the Higgs fields at the IR (TeV) brane
but its decay constant is given by a mass scale hierarchically larger than the TeV scale.
The presence of the intermediate brane makes it possible while solving the electroweak naturalness problem.

The axion mass generated from non-perturbative QCD effects is given by the usual formula,
\begin{align}
\label{eq:axion_mass_QCD}
m_a^{\rm QCD}= \frac{\sqrt{z}}{1+z}\frac{m_\pi F_\pi}{F_a}\ ,
\end{align}
where $z$ is the ratio of the up-quark and down-quark masses, $z \equiv {m_u}/{m_d}\simeq 0.56$,
and $m_\pi, F_\pi$ denote the pion mass and decay constant, respectively.
To estimate the axion quality, we now define the quality factor as $\mathcal{Q}\equiv(m_a^{\cancel{\rm PQ}}/m_a^{\rm QCD})^2$.
The strong CP problem is correctly solved with $\mathcal{Q} \leq 10^{-10}$.
Fig.~\ref{fig:quality} shows the contours of the quality factor $\mathcal{Q}$ as a function of $\Delta_1$
for different values of the axion decay constant $F_a$.
We can see that $\mathcal{Q}$ is significantly suppressed as $\Delta_1$ increases
and the axion quality problem is solved for $\Delta_1 \gtrsim 10$.
A larger $F_a$ needs a larger $\Delta_1$ to address the problem.

\section{Predictions}
\label{KK_modes}

The model shows several characteristic features which do not exist in the ordinary DFSZ axion model
or its realization in the two 3-brane setup \cite{Cox:2019rro}.
First, it predicts the KK towers of the axion, $\eta$ (the radial mode of $\Phi$) and the $U(1)_{\rm PQ}$ gauge field 
at around the typical mass scale of the IR brane,
in addition to the KK towers of the SM gauge bosons and radions corresponding to the intervals between the branes
(the radion masses have been presented in ref.~\cite{Lee:2021wau}).
The mass spectrum of the KK axions and their couplings to the SM fields are analyzed.
We briefly discuss collider phenomenology and cosmology of the KK axions but their detailed studies are left for a future study.
The model also predicts relatively light extra Higgs bosons in the two doublet Higgs fields $H_{u,d}$
whose discoveries are expected at the Large Hadron Collider (LHC) or future collider programs.

\subsection{KK axions}
\begin{figure}
\centering
     \includegraphics[width=0.45\linewidth]{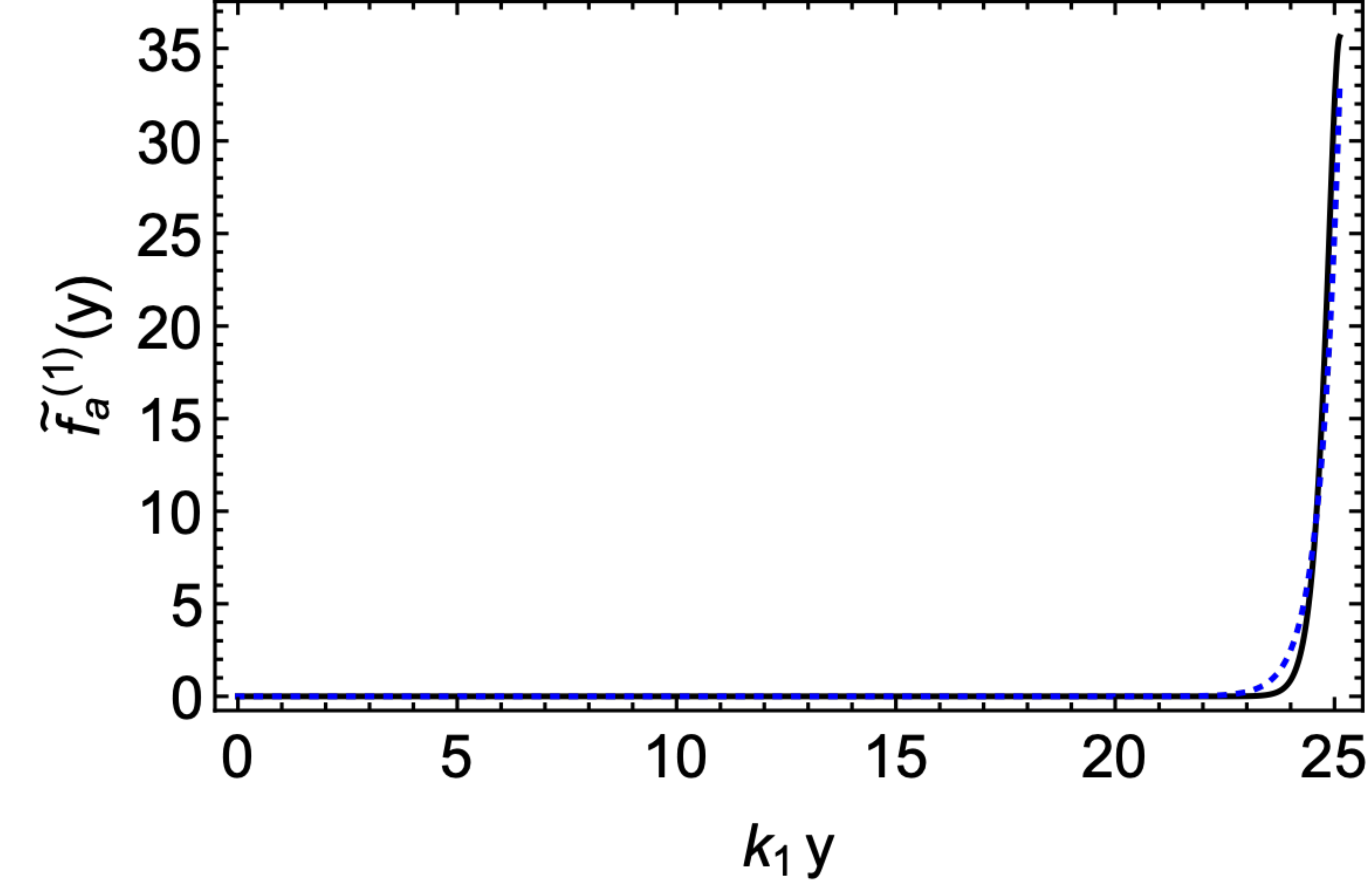}\\
   \caption{
The numerical (solid black) and approximated analytic (blue dashed) results for the bulk profile of the first KK excitation mode
$\tilde f_a^{(1)}(y) \equiv k_1e^{-k_2(y_{\rm IR}-y_{\rm I})-k_1y_{\rm I}}\,f_a^{(1)}(y)$.
 We take $y_{\rm IR}=\ln(10^{12})/k_2$, and the other parameters are the same as those used in Fig.~\ref{fig:background}.
    For the solution of $\langle\Phi\rangle$, we have used the simplified formula in Eq.~\eqref{eq:c_abcd}.
   }
\label{fig:ma_KK}
\end{figure}

For the KK excitations of the axion $a$,
the solutions for the bulk equations have been obtained in Eq.~\eqref{eq:a1}.
The constants $C_{a1}, C_{a4}$ are determined in terms of $C_{a2}$ and $C_{a3}$
by using the boundary conditions at the UV and IR branes in Eqs.~\eqref{aUVcondition}, \eqref{aIRcondition}.
Besides, the constant $C_{a3}$ is fixed by the continuity of $a$ at the intermediate brane in Eq.~\eqref{continuitya},
and $C_{a2}$ is the overall factor.
The KK mass spectrum is obtained by investigating $m_a$ to satisfy the last boundary condition of the continuity of $\partial_5 a$ in Eq.~\eqref{continuitya}.
For the parameter set taken in Fig.~\ref{fig:background},
the mass of the first KK excitation mode is approximately given by
$m_a^{(1)} \approx 4 k_2e^{-k_2y_{\rm IR}-(k_1-k_2)y_{\rm I}}$ that is around the typical mass scale of the IR brane.
Fig.~\ref{fig:ma_KK} presents the bulk profile of the first KK excitation mode.
We can see that the profile is localized toward the IR brane.
Around the IR brane, the profile  is approximated as $f^{(1)}_a(y) \approx
e^{k_1y_{\rm I}-k_2(y_{\rm IR}+y_{\rm I}(\Delta_2-3))}e^{k_2y(\Delta_2-2)}/(\sigma_0\sqrt{k_1k_2})$
which agrees well to the numerical result in the figure.

Like the lightest mode, the KK axions couple to the SM fields,
but interestingly their interaction strength is much larger than that of the lightest mode of the axion.
For example, the coupling of the first KK excitation mode to the gluons is given by
\begin{align}
\begin{split}
&S_{\rm eff} \supset \int d^4 x\,\frac{1}{32\pi^2 F_{a^{(1)}_{KK}}}a^{(1)}(x)G^{a(0)}_{\mu\nu} \widetilde G^{a(0)\mu\nu} \ ,
\qquad
\frac{1}{F_{a^{(1)}_{KK}}} \equiv A_{\rm QCD}\,f^{(1)}_a(y_{\rm IR}) \ .
\end{split}
\end{align}
The decay constant of the first KK excitation mode scales as $F_{a^{(1)}_{KK}} \sim k_2\,e^{-k_2(y_{\rm IR}-y_{\rm I})-k_1y_{\rm I}}$
for $\Delta_2\approx 4$,
$i.e.$ the first KK axion couples to the gluons or the other SM fields with the decay constant at around the typical scale of the IR brane
while the decay constant of the lightest mode is determined by the much larger scale of the intermediate brane.
Therefore, the KK axions are visible while our model realizes an invisible axion.
Searches for such KK axions have been conducted in the context of the so-called axion-like particles.
In the present model, the KK axions are expected to be at around the TeV scale so that they could be a target of the LHC
and future collider experiments~\cite{Jaeckel:2012yz,ATLAS:2017ayi,Molinaro:2017rpe,Baldenegro:2018hng,Bauer:2018uxu,Inan:2020kif,dEnterria:2021ljz}.
At colliders with sufficiently high energies, the KK axions are resonantly produced through $e.g.$ the gluon-gluon fusion
and their decays show diphoton resonances.

In the early Universe, the KK axions are produced through the sizable couplings to the SM fields, but
once the Universe cools down below the temperature of their masses, they quickly decay into the SM particles
and do not cause any cosmological issues.
The cosmological evolution of the lightest mode of the axion follows the standard story and it provides a dark matter candidate.

In addition to the KK axions, there exist KK excitation modes for $\eta$ and the $U(1)_{\rm PQ}$ gauge field.
Their first excitation modes have masses at around the typical mass scale of the IR brane
$\sim k_2\,e^{-k_2(y_{\rm IR}-y_{\rm I})-k_1y_{\rm I}}$.
The KK modes of $\eta$  couple to the SM fields through Eq.~\eqref{eq:Phi2HuHd},
and the KK $U(1)_{\rm PQ}$ gauge bosons couple to the SM particles with nonzero $U(1)_{\rm PQ}$ charges.
As in the case of the KK axions, they can be produced at colliders and cosmologically harmless.

\subsection{Extra Higgs bosons}
\label{sec:pheno}

\begin{figure}
\vspace{-0.2cm}
\centering
   \includegraphics[width=0.43\linewidth]{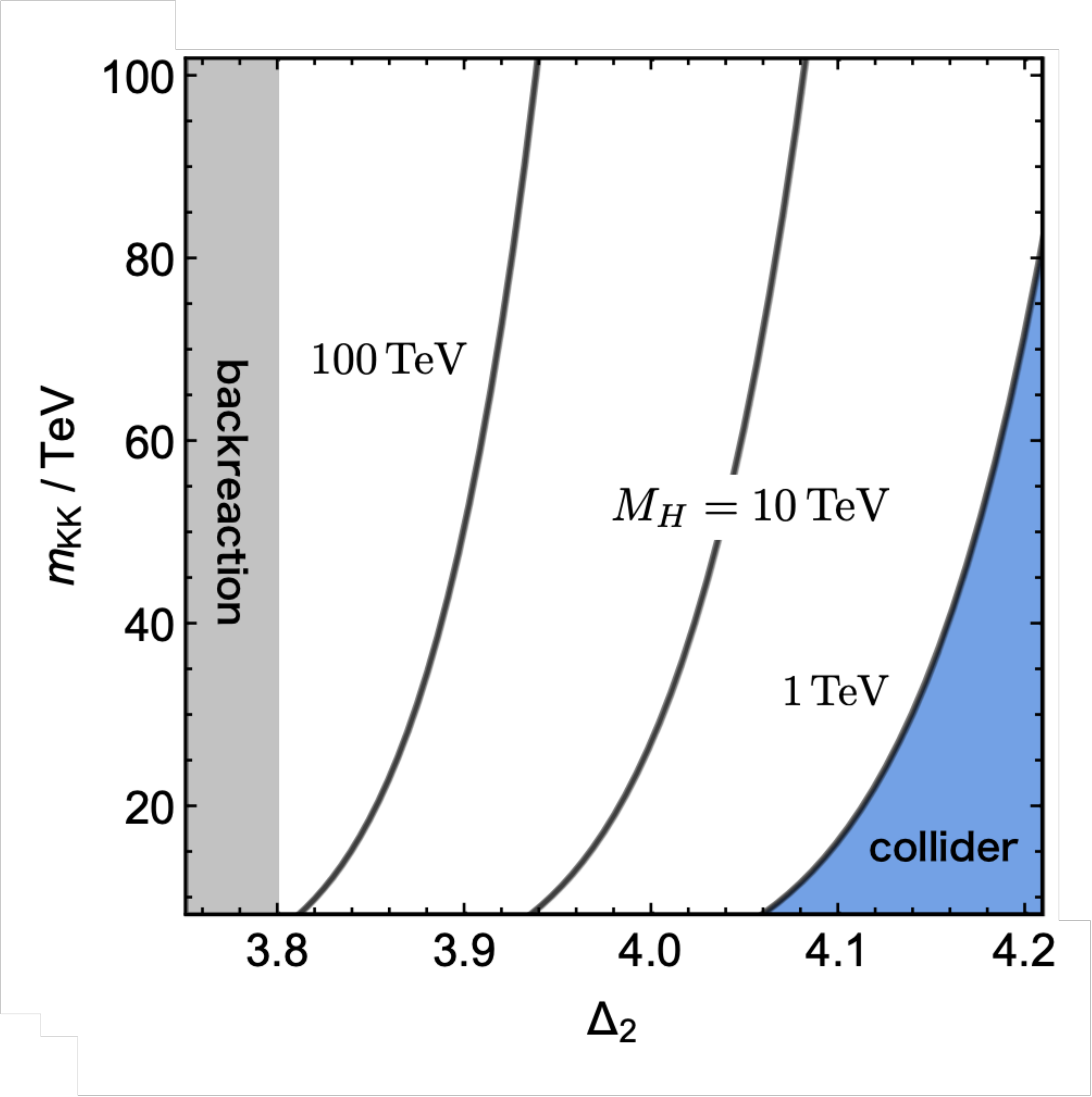}
   \vspace{-0.1cm}
   \caption{The mass of the extra Higgs bosons $M_H$ as a function of $\Delta_2$
and the typical mass of the first KK excitation $m_{KK}\equiv 4k_2\,e^{-k_2(y_{\rm IR}-y_{\rm I})-k_1y_{\rm I}}$.
Here, we take $\kappa_{ud}=1, \sigma_0=1, \tan\beta=10, k_{1,2}=0.1M_{\rm Pl}$ and $y_{\rm I}=\ln(10^6)/k_1$.
The blue shaded region gives $m_H\leq 1\,{\rm TeV}$ which is constrained at collider experiments.
In the gray shaded region, the backreaction of the PQ field to the metric is not negligible.}
\label{fig:mH}
\end{figure}

The two $SU(2)_L$ doublet Higgs fields $H_{u}, H_d$ live on the IR brane and couple to the PQ field $\Phi$ through
the term in Eq.~\eqref{eq:Phi2HuHd}.
A sizable $\langle \Phi (y_{\rm IR}) \rangle$ is needed to make the extra Higgs bosons sufficiently heavy.
In fact, when we assume the two-Higgs-doublet model of type-II, the mass of the extra Higgs bosons is approximately given by
\begin{align}
M_H^2\approx\frac{2\,e^{-2k_2\,y_{\rm IR}-2(k_1-k_2)\,y_{\rm I}}\,\kappa_{ud}\,\langle\Phi (y_{\rm IR}) \rangle^2}{\sin(2\beta)M_5}\ ,
\end{align}
where $\tan \beta \equiv \langle H_u\rangle/\langle H_d\rangle$ and we have focused on the region of $M_H\gg 100\,{\rm GeV}$.
The computation in Sec.~\ref{sec:background} gives
$\langle\Phi (y_{\rm IR}) \rangle \approx k_1^{3/2}\sigma_0 e^{-k_2(y_{\rm IR}-y_{\rm I})(\Delta_2-4)}$.
It indicates that $\langle\Phi (y_{\rm IR}) \rangle$ receives an exponential suppression for $\Delta_2> 4$,
which leads to relatively light extra Higgs bosons compared to the typical mass scale of the IR brane.
Fig.~\ref{fig:mH} shows the mass of the extra Higgs bosons $M_H$ as a function of $\Delta_2$
and the typical mass of the first KK excitation $m_{KK}\equiv 4k_2\,e^{-k_2(y_{\rm IR}-y_{\rm I})-k_1y_{\rm I}}$.
For $\Delta_2 < 4-\ln(6M_5^3/\sigma_0^2k_1^3)/(2k_2(y_{\rm IR}-y_{\rm I}))$ (gray shaded region),
the backreaction of the PQ field to the metric is not negligible.
We put a reference collider constraint which excludes $M_H \leq 1\,{\rm TeV}$ (blue shaded region). 
The mass region of $M_H\gtrsim1\,{\rm TeV}$
can be probed at the LHC and future collider experiments
\cite{Hajer:2015gka,Craig:2016ygr,Gu:2017ckc,Chen:2018shg,Kling:2018xud,Li:2020hao,Han:2020lta,Han:2021udl}.
See also $e.g.$ refs.~\cite{Deschamps:2009rh,Arnan:2017lxi,Atkinson:2021eox} for further analyses
of searching for the viable parameter space of two Higgs doublet models.

\section{Discussions}\label{discussions}

In order to concurrently address the electroweak naturalness problem
and a high-quality axion problem, we study a novel mechanism based on a doubly composite dynamics where the second confinement takes place after the CFT encounters the first confinement and the theory flows into another conformal fixed point. In particular, via AdS/CFT, 
we presented our work in the warped extra dimension model with three 3-branes.
The typical mass scales of the UV, intermediate and IR branes are identified as the Planck, spontaneous $U(1)_{\rm PQ}$ breaking and TeV scales.
The two doublet Higgs fields live on the TeV brane so that the naturalness problem is addressed as in the original RS model.
The $U(1)_{\rm PQ}$ symmetry is realized as a gauge symmetry and is only spontaneously broken in the whole space except for the UV brane.
We have introduced a 5D scalar field whose potential at the intermediate brane drives a spontaneous breaking of the $U(1)_{\rm PQ}$ symmetry.
The profile of the lightest mode of the axion is significantly suppressed around the UV brane, which protects the axion from
gravitational violations of the $U(1)_{\rm PQ}$ symmetry on the UV brane.
The $U(1)_{\rm PQ}$ gauge anomaly is correctly canceled by including the 5D Chern-Simons term.
The low-energy effective theory contains the axion coupling to the Higgs fields
and reduces to the DFSZ axion model that respects the approximate $U(1)_{\rm PQ}$ global symmetry.
We have shown that the strong CP problem is correctly solved by a high-quality axion.
The marriage of two different solutions for the electroweak naturalness and the axion quality problem in a single theory framework
provides a novel phenomenology relevant for future experiments. 
The KK towers of the axion, the radial mode of the PQ field and the $U(1)_{\rm PQ}$ gauge field exist
at around the typical mass scale of the IR (TeV) brane.
Since the first KK axion has a decay constant at around the TeV scale, its discovery at future colliders
will be a smoking gun evidence for our model.
Light extra Higgs bosons are also likely to exist when the VEV of the PQ field is suppressed at the IR brane,
and their discoveries are expected at the LHC or future collider experiments.

In the present work, we have assumed that the SM quarks and leptons are all localized at the IR brane for simplicity.
However, it has been well known that the bulk fermions with different profiles can naturally realize the SM flavor structure
\cite{Grossman:1999ra,Chang:1999nh,Gherghetta:2000qt,Huber:2000ie,Agashe:2004cp}.
In this case, KK gluons contribute to flavor-changing neutral current processes
which puts a lower bound on their masses at $\mathcal{O}(10) \, \rm TeV$ (see $e.g.$ ref.~\cite{Csaki:2008eh}).
In our model, the assignment of the $U(1)_{\rm PQ}$ charge for each SM fermion is another ingredient that may generate a rich flavor physics.
A flavor-dependent assignment leads to new contributions to flavor observables which might be useful to probe our model.
Conversely, it might be interesting if the $U(1)_{\rm PQ}$ symmetry could suppress dangerous contributions to flavor observables and
help to lower the mass scale of KK excitation modes.

A multiple 3-brane model predicts 
multiple confinement-deconfinement phase transitions in the early Universe
\cite{Creminelli:2001th,vonHarling:2017yew,Baratella:2018pxi,Fujikura:2019oyi}.
They are typically of the strong first order and generate
detectable gravitational waves (GWs) with multiple peak frequencies.
In our model, one frequency is likely to be within the range covered by LIGO
and the other may be also within the range of future space-based GW observers.
Furthermore, the new first order phase transition near the TeV scale might be able to realize electroweak baryogenesis
\cite{Bruggisser:2018mus,Bruggisser:2018mrt}.
The model with a new CP phase needed for baryogenesis is then likely to be probed by the electron EDM.

\section*{Acknowledgements}
We would like to thank Peter Cox for useful discussions. 
S.L.\ was supported by the National Research Foundation of Korea (NRF) grant funded by the Korea government (MEST) (No. NRF-2021R1A2C1005615).
S.L.\ was also supported by the Visiting Professorship at Korea Institute for Advanced Study.

\bibliographystyle{utphys}
\bibliography{bib}

\end{document}